%% file: main.tex
\documentclass[sigplan,screen]{acmart}
\usepackage{xcolor}
\usepackage[normalem]{ulem}
\usepackage{enumitem}

\AtBeginDocument{%
  }

\setcopyright{acmlicensed}
\copyrightyear{2026}
\acmYear{2026}
\acmConference[FARM '26]{Functional Art, Music, Modeling and Design 2026}{August 24th, 2026}{Indianapolis, IN}

\begin{document}

\title{Composition: Building Community with Arts, Math, and Code (Experience Report)}

\author{Isidore Mohr}
\authornote{Both authors contributed equally to this research.}
\email{isidoremohr@gmail.com}
\orcid{0009-0008-5038-028X}
\affiliation{
    \city{Philadelphia}
    \state{Pennsylvania}
    \country{USA}
}

\author{Claire Wang}
\authornotemark[1]
\email{cdwang@engineering.upenn.edu}
\orcid{0009-0002-9321-0877}
\affiliation{%
  \institution{University of Pennsylvania}
  \city{Philadelphia}
  \state{Pennsylvania}
  \country{USA}
}
\renewcommand{\shortauthors}{Mohr and Wang}

\begin{abstract}
  We report our experiences co-organizing Composition, a free event series on art, mathematics, and code. We discuss the structure of the event, our process for selecting artists, outreach efforts for submissions and event promotion, and the community response.
\end{abstract}

\maketitle

\input{parts/introduction}
\input{parts/related}
\input{parts/cfp}
\input{parts/venues}
\input{parts/event}
\input{parts/outreach}
\input{parts/community}
\input{parts/conclusion}

\begin{acks}
We are grateful to everyone who has presented at Composition: Micah Fitch, Brian Abelson, Michael Wehar, Matthew Kaney, Matt Zucker, arjun, and Nick Romano. We thank Wonderville for the use of their space and their help, and Nicole Harwayne for providing audio support at Wonderville. We are also grateful to The Rotunda for the use of their space, and their staff for all their help: Gina Renzi, Gladys Nobriga (audio support), Styx Latté (audio support), and The Rotunda's student volunteers. Finally, we thank Zachary Johnson and Jp Smith for volunteering at our events.
\end{acks}

\bibliographystyle{ACM-Reference-Format}
\bibliography{main}

\appendix
\include{appendix}

\end{document}

%% file: parts/introduction.tex
\section{Introduction}
\label{sec:intro}


Composition \cite{composition_website} is a bi-annual free event series on the intersection of arts, math, and code, located in the Northeast region of the United States. The event serves as a meetup and community platform for anyone to present their work, as long as it 1) relates to arts and math or code, and 2) demonstrates technical depth. Work in progress is welcome.

So far, Composition has had two events: Composition \#0 \cite{noauthor_composition_2025}, held at Wonderville \citep{noauthor_wonderville_nodate} in Brooklyn, NY in December 2025; and Composition \#1 \cite{noauthor_composition_nodate}, held at The Rotunda \citep{noauthor_rotunda_nodate} in Philadelphia, PA in May 2026.

We present an experience report on all facets of our event as a case study and reference point for building community centered around arts, math, and code. The report shares:
\begin{itemize}[noitemsep,nolistsep]
\item Our approach for selecting speakers via a submission process.
\item Outreach efforts for finding submissions and promoting the event.
\item How each Composition event is structured.
\item Observations about the community around Composition, featuring feedback from attendees.
\end{itemize}

%% file: parts/related.tex
\section{Related}

Composition's main inspirations are the Functional Art, Music, Modeling, and Design Workshop (FARM) \citep{noauthor_workshop_nodate}, livecoding events, and regional academic events such as the New Jersey Programming Languages and Systems Seminar (NJPLS) \citep{noauthor_nj_nodate}. FARM's mixture of theoretical academic work, fun projects, and live performances is a compelling model, which Composition intends to replicate and build on. NJPLS is a free bi-annual tri-state region event for the programming languages community, often hosted at venues accessible by public transit. Algoraves are music events that explore algorithms with live electronic dance music, often featuring live coding to compose music \citep{collins2014algorave}. Composition aims to provide an experience combining these qualities of FARM, NJPLS, and algoraves at a local and accessible scale.

There are many other events dealing with the relationship between art, mathematics, and technology. Bridges is a conference on mathematical connections in art, music, architecture, and culture \citep{noauthor_bridges_nodate}. Matt Zucker \citep{noauthor_matt_nodate} and Michael Wehar \citep{noauthor_michael_nodate}, two of the speakers at Composition \#1, have previously presented at Bridges, which is how we learned of their work. The International Conference on Computer Music \citep{noauthor_icmc_nodate} brings together artists working in all areas of computer music, sound art, and multimedia. The American Mathematical Society's Joint Mathematics Meetings \citep{noauthor_ams_nodate} feature mathematical art, including an earlier showing of Matt Zucker's Penrose tiling quilts. 

%% file: parts/cfp.tex
\section{Approach to Submissions}
\label{sec:submissions}
\subsection{Philosophy}
Composition's overarching goal is to be accessible and worth accessing, for both the audience and the artists. It does not require proposals to be polished, and gladly accepts works in progress. The criterion is that submissions should have artistic value to audiences with or without the relevant mathematical or computational background. Art can make unfamiliar concepts interesting to audiences who would otherwise be lost and confused, and technical depth rewards the audience's engagement. Variety of art modalities is also a goal. Composition \#0 had primarily musical works, though some had accompanying visuals. For Composition \#1, we selected a mix of purely visual, primarily musical, and audiovisual works.

\subsection{Method}
\label{sec:submissions:method}
\subsubsection{Submission Form.} We receive proposals through a Google form. Composition \#0 was very time-limited, so the form was only open for two weeks before we had to finalize the lineup. For Composition \#1, we had a three month submission period, keeping the submission form open until a few days before we had to inform the venue of the lineup. The extended period helped us gather several more submissions. We could have benefited by moving the deadline slightly earlier, so that we could have more time to evaluate submissions. See Appendix ~\ref{app:submissions} for the submission form questions.

\subsubsection{Outreach for Submissions.} We spread our call for submissions to as wide a range of communities as possible. Groups that are more directly related to Composition have a higher density of potential submitters, but any group may have interested members. We make use of the Composition website, social media, mailing lists, and Discord servers for events, art, and technical communities. Such communities include Livecode NYC \citep{noauthor_livecodenyc_nodate}, the Philadelphia ACM SIGGRAPH Chapter \citep{noauthor_siggraph_nodate}, and local event groups. For these broad platforms we post the call for submissions as a flyer and as text, including the time and place of the event as well as links to the submission form, our website, and our social media.
\subsubsection{Sourcing Submissions Individually.} We also individually contact artists, taking a more focused approach. From Composition \#1 onward, in addition to artists with whom we have existing connections, we search through other venues that are likely to host relevant artists. These may be events on art-technical crossovers, such as the Bridges conference, the International Conference on Computer Music, and the art presented at American Mathematical Society's Joint Mathematics Meetings. We also look at local art communities, such as Philadelphia's InLiquid \citep{noauthor_inliquid_nodate}. We try to reach out to a large and diverse group of artists from among these communities. We send artists the event information and our links along with a mention of how we found their work and what we were interested in.

\subsection{Results}
The form for Composition \#0 received four submissions, not counting the organizers. We accepted three of these submissions, from two submitters. The form for Composition \#1 received nine responses, of which eight were submissions. We accepted six of these submissions; one submitter had scheduling conflicts and dropped out.

%% file: parts/venues.tex
\section{Venues}
To find venues, we look at local spaces for art events. Since Composition has a variety of modalities, the best spaces host multiple types of events, including but not limited to music. They should accommodate free events, and have reasonable fees. Ideally, they should also be near public transit.
Composition \#0 was held at Wonderville, a bar and DIY community space that is popular with the livecoding community \citep{noauthor_wonderville_nodate}. We reached out to them about two months in advance, which was enough to book a date but left us very little time to finalize our lineup. Composition \#1 was held at the Rotunda, a community event space in West Philadelphia \citep{noauthor_rotunda_nodate}. The Rotunda is owned by the University of Pennsylvania, but its programming is mostly independent, and includes a diverse collection of art and music events. We reached out about four and a half months in advance, which gave us plenty of time to resolve the details. In this case, we were lucky to find an available date, as the Rotunda usually fills up weekends six or more months ahead. Both venues were excellent spaces with good sound and visual support.

%% file: parts/event.tex
\section{Event}
\subsection{Event Structure}
Composition consists of a series of presenters on different topics connecting art with mathematics or code. Presentations do not have a predefined format: slides, live demos of tools, performances, and physical art objects have all been part of Composition. Each presentation is followed by a question and answer session. Events are free, with optional donations accepted to support the venues.

The duration of the event is \textasciitilde 3 hours. This time includes 4 or 5 speakers at roughly 20-40 minutes each, with 5 minutes each for Q\&A, and additional time of at least 20 minutes for the event as a whole. Composition \#0 divided time evenly between presenters, while for Composition \#1 we gave each presenter as much time as they requested. The Q\&A doubles as setup time for the next presenter. The extra time serves as a buffer for talks and questions that go past their appointed time, and gives attendees a chance to chat afterwards. See Appendix ~\ref{app:event} for pictures from previous Composition events.

\subsection{Audio Visual Support}
The event is recorded, with recordings \citep{composition_youtube} of each speaker put on YouTube afterwards with their consent. At Composition \#0, the venue had a preexisting recording setup; at Composition \#1, we borrowed a video camera from Penn Libraries \citep{noauthor_penn_nodate}. Audience members were inaudible in recordings of Q\&A sessions at the first event, as they were not connected to the venue's sound system. This was fixed for the second event by having one of the cohosts pass a microphone between audience members.

\subsection{Audience Participation}
\subsubsection{Q\&A} The Q\& A after each presentation consists of audience members asking questions, with one of the cohosts identifying members and providing the audience a microphone.
\subsubsection{Audience Conversation with a Presenter.} In Composition \#0, Micah Fitch one of the presenters, additionally submitted a ``conversation'' proposal on the relationship between audio sequencers and compilers. This was intended to be a more interactive session with less of a divide between presenter and spectators; unfortunately, the audience largely disintegrated. This was likely in part because in part because audience members found the question difficult and in part because the event was nearing its end. A lack of structure may also have made it unclear how to engage. Future events could experiment with conversational sessions, but would need further effort to ensure the whole audience can meaningfully participate.

\subsection{Survey of Performances at Composition}
In this section, we list and summarize the presentations at Composition \#0 and Composition \#1. 

\subsubsection{Composition \#0}
 \hfill\\
    \textbf{Isidore Mohr -- Girard's Paradox.} Discussion and performance of a musical piece based on a proof term for Girard's paradox \citep{mohr_demo_2025}. \\
    \textbf{Claire Wang -- Music, Proofs, and Dependent Types.} Discussion and demo of two projects: Rocq N' Roll \citep{burtonpatel_demo_2025}, which turns tactics in the Rocq Prover into music, and Prismriver \citep{aniva_prismriver_nodate}, a domain-specific language for formalizing music theory such as counterpoint using the proof assistant Lean. \\
    \textbf{Micah Fitch \citep{noauthor_front_nodate} -- Polyphonotopes.} Demo of a theoretical framework and tools for mapping and understanding relationships between pitch-class sets.\\
    \textbf{Brian Abelson -- saysynth.} Discussion of the history of speaking machines and demo of Saysynth \citep{noauthor_saysynth_nodate}, a synthesizer built on top of Mac's text-to-speech framework.\\
    \textbf{Micah Fitch -- Conversation: ``Compiling'' to interactive dynamic sequencers?} A conversation among the audience, led by Micah Fitch, on the relationship between computer programs, interactive sequencers, and compilers.
\subsubsection{Composition \#1}
\hfill\\
    \textbf{Michael Wehar -- Algoart.} Presentation and demo of a web-based platform, Algoart \citep{noauthor_algoart_nodate}, that brings together coding and visual arts.\\
    \textbf{Matthew Kaney - Composing Patterns.} Presentation and demo of the theory \citep{mclean_uzulangs_2026} behind ``uzulangs,'' the family of live-coding pattern tools based on TidalCycles and Strudel.\\
    \textbf{Matt Zucker  -- Computer-Assisted Quilting.} Discussion and presentation of three quilts based on mathematical tessellations, along with the software and fabrication techniques used to produce them \citep{noauthor_penrose_2022,noauthor_pentagonal_2022}.\\
    \textbf{arjun \citep{noauthor_arjuns_nodate} -- algorithms for expression \& truth.} Presentation of several art pieces whose common thread is using logic-systems to understand lived-human-experiences, and building algorithms to express them.\\
    \textbf{Nick Romano \citep{noauthor_nick_nodate} -- Techno-juggling.} Demo of a color-tracking juggling table with discussion of historical contacts between juggling and technology.\\

%% file: parts/outreach.tex
\section{Outreach and Promotion}
\label{sec:outreach}
\subsection{Social Media Presence}
Composition has a persistent online presence in its website \cite{composition_website}, on social media, and YouTube. This presence enables interested parties to learn about Composition, keep up to date on events, and share events by word of mouth. Composition's social media presence is primarily on Instagram \cite{composition_instagram}, because of its large communities around art, including technical art. Composition's YouTube channel hosts event recordings.

\subsection{Promotional Materials}
Once the lineup for the event is settled, we create promotional materials, and an event page for sharing and tracking RSVPs. Flyers are minimal and not optimized for marketability, but they convey event details succinctly. RSVPs are useful to estimate how many people will attend, but attendees have outnumbered RSVPs: Composition \#1 received 28 RSVPs, including RSVPs made the day of the event, and at least 38 people attended. See Appendix ~\ref{app:graphics} for examples of promotional materials from previous Composition events.
\subsection{Event Outreach Process}
We begin to spread the word about Composition about a month in advance, using the same broad platforms mentioned in section \ref{sec:submissions:method}, with the addition of websites, mailing lists, and Signal chats for local event promotions. Venues' social media presences are especially important in this phase, as they already have an audience interested in the events they host. We post event information on most platforms once or twice during the month leading up to the event, though on Composition's own social media we may post more frequently to remind potential attendees about the event.

%% file: parts/community.tex
\section{Community}
\label{sec:community}
Both instances of Composition have had around 35-40 attendees -- a substantial turnout for a new event. Audiences have been highly engaged, usually asking presenters multiple thoughtful questions. In some cases, when Q\&As began with silence, cohosts broke the silence by asking questions which sparked lively follow-ups from the audience.

At both events, audience members and venue staff have mentioned that Composition is an unusual sort of event that they had not seen before at the venue, but which they appreciated. As an academic-style event that is accessible to a broad audience in cost, location, and content, Composition serves a mostly unfilled niche. Composition can introduce people to new technical concepts. For example, two of the presenters at Composition \#0 discussed proof assistants. Attendees at Composition \#0 expressed gratitude for exposure to topics they didn't realize existed.

\subsection{Feedback and Testimonials}
Composition has an on-going public form for submitting feedback. In this section we share testimonials from event attendees. See Appendix ~\ref{app:feedback} for the feedback form questions.

\textbf{Novelty.} Multiple attendees expressed the novelty of the event due to the topics and the format of Composition.
\begin{itemize}
\item \itshape ``This was my first time experiencing live performances of algo/math-art.''
  \item \itshape ``I really enjoyed it! As a non-tech person I didn't understand the math/algorithmic/programming side of things, but they were still interesting projects and were presented in a way that I could still follow along.''
  \item \itshape ``I appreciated being able to hear from the artists themselves, about their process, which is not always available on the internet.''
\end{itemize}
\textbf{Variety of Content.} Attendees from Composition \#1 noticed the variety of art forms amongst the presenters. 
\begin{itemize}
    \item \itshape ``..and the variety of projects. I liked the creative projects as they showed how programming and code can be implemented in less abstract, less academic contexts (quilting, visual art, music, etc).''
    \item \itshape ``Fantastic variety in both the art forms and how the artists intertwined computer science and art.'' 
\end{itemize}
\textbf{Community.} Attendees found community with each other throughout a Composition event. 
\begin{itemize}
\item \itshape ``The audience also seemed to come from a lot of different backgrounds.''
\item \itshape ``You had padding time that allowed both for some flexibility in how long the presentations ran and for socializing after.''
\end{itemize}
\textbf{Inspiration.} At least one attendee felt inspired to create their own projects after attending Composition.
\begin{itemize}
\item \itshape ``Their passion was infectious, and I was motivated to try quilting on my own after discussing an idea I had with the artist about a pleasing geometric construction arising from 2D Sturmian sequences.''
\end{itemize}

%% file: parts/conclusion.tex
\section{Conclusion}
Composition is a regional meetup on arts, math, and code where featured speakers present complete work and works in progress that demonstrate technical depth entwined with art. The experience report shares the entire process of how a Composition event is produced, a survey of presenters at Composition, and feedback from the community. Composition strives to continue to organize with the intention of building accessibility and connection to arts, math, and code to its surrounding local communities. 



%% file: appendix.tex
\input{appendix/Event}

\newpage

\input{appendix/Graphics}

\newpage

\input{appendix/CFP}

\newpage

\input{appendix/Feedback}

%% file: appendix/Event.tex
\section{Composition Event}
\label{app:event}
The following are images from Composition's events.


\begin{figure}[H]
\centering
\includegraphics[width=\linewidth]
{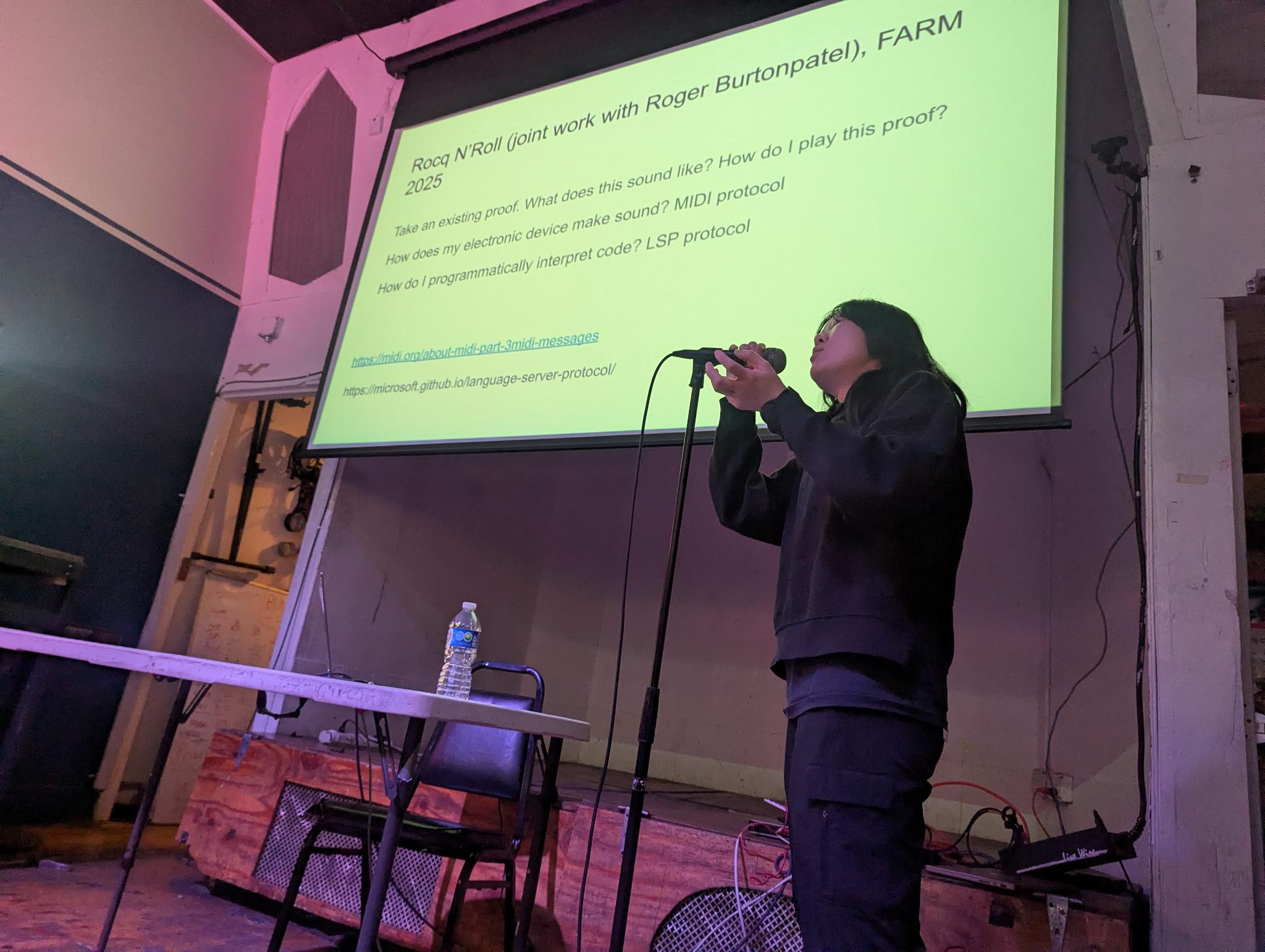}
\caption{Claire Wang presenting at Composition \#0.}
\Description{Talking into mic in front of a projection of dotted circles and lines. TODO}
\label{fig:appendix-composition-event-clairewang}
\end{figure}

\begin{figure}[H]
\centering
\includegraphics[width=\linewidth]
{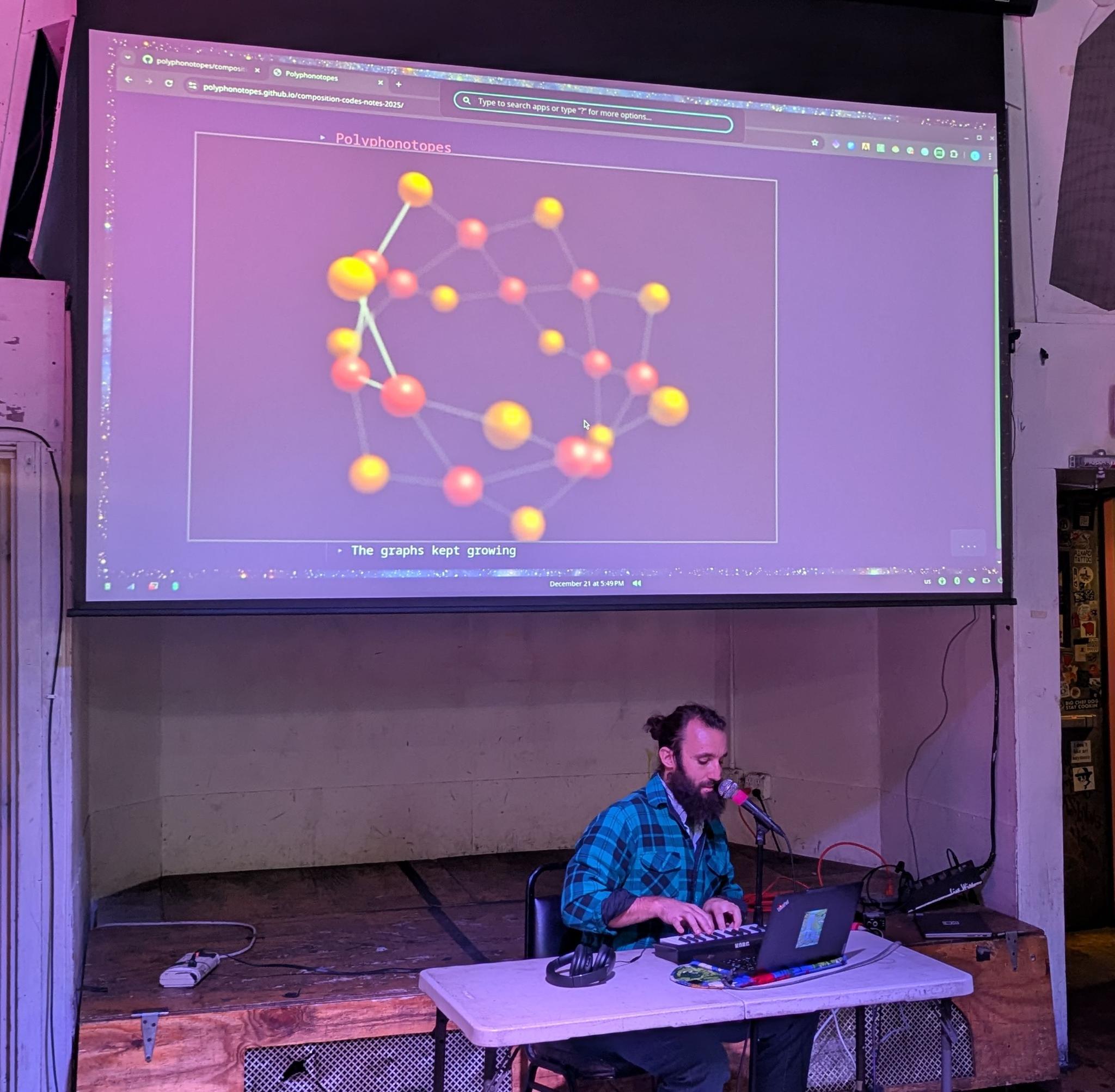}
\caption{Micah Fitch presenting at Composition \#0.}
\Description{Talking into mic in front of a projection of dotted circles and lines. TODO}
\label{fig:appendix-composition-event-micahfitch}
\end{figure}

\begin{figure}[H]
\centering
\includegraphics[width=\linewidth]
{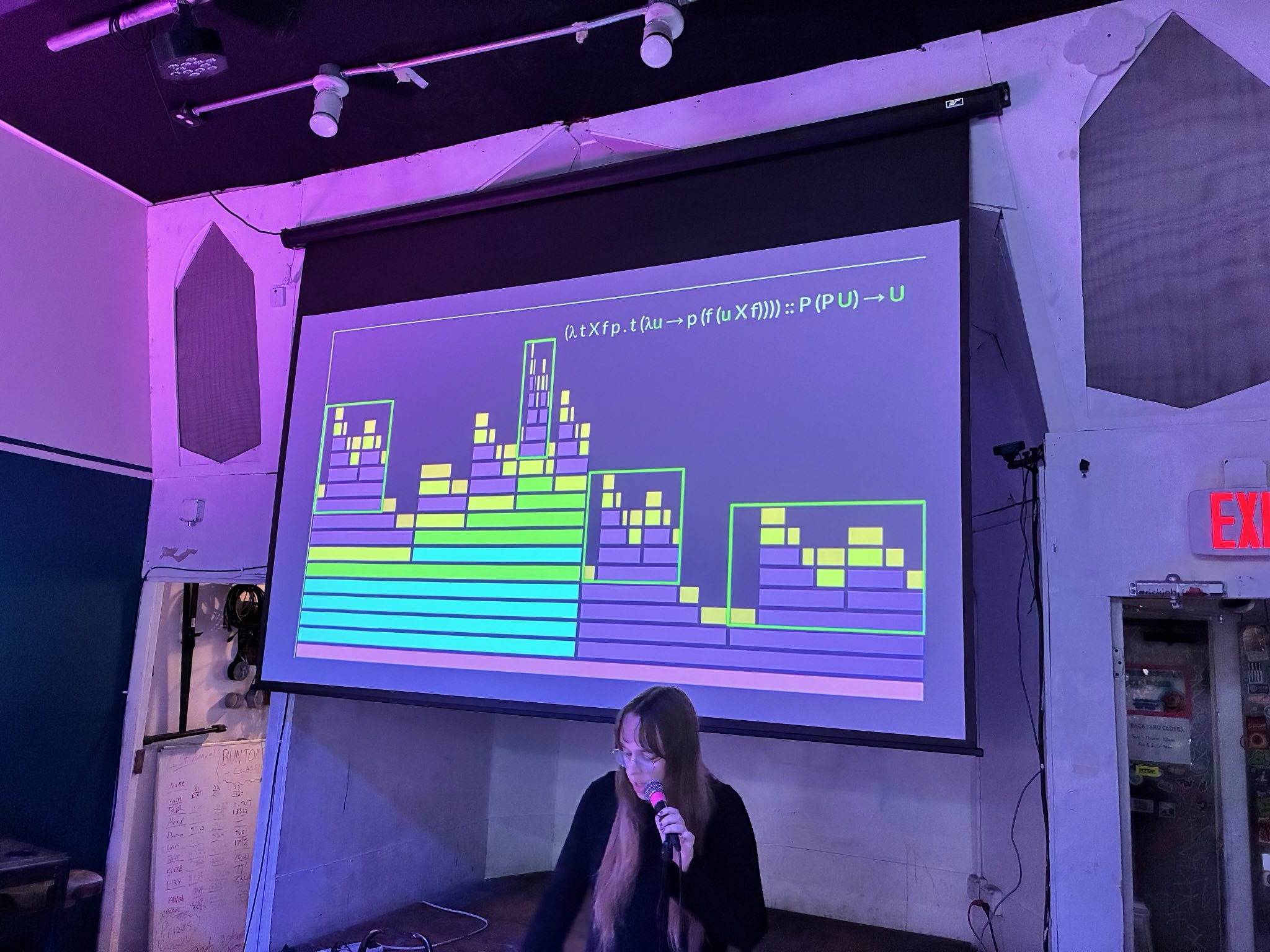}
\caption{Isidore Mohr presenting at Composition \#0.}
\Description{Talking into mic in front of a projection of dotted circles and lines. TODO}
\label{fig:appendix-composition-event-isidoremohr}
\end{figure}

\begin{figure}[H]
\centering
\includegraphics[width=\linewidth]{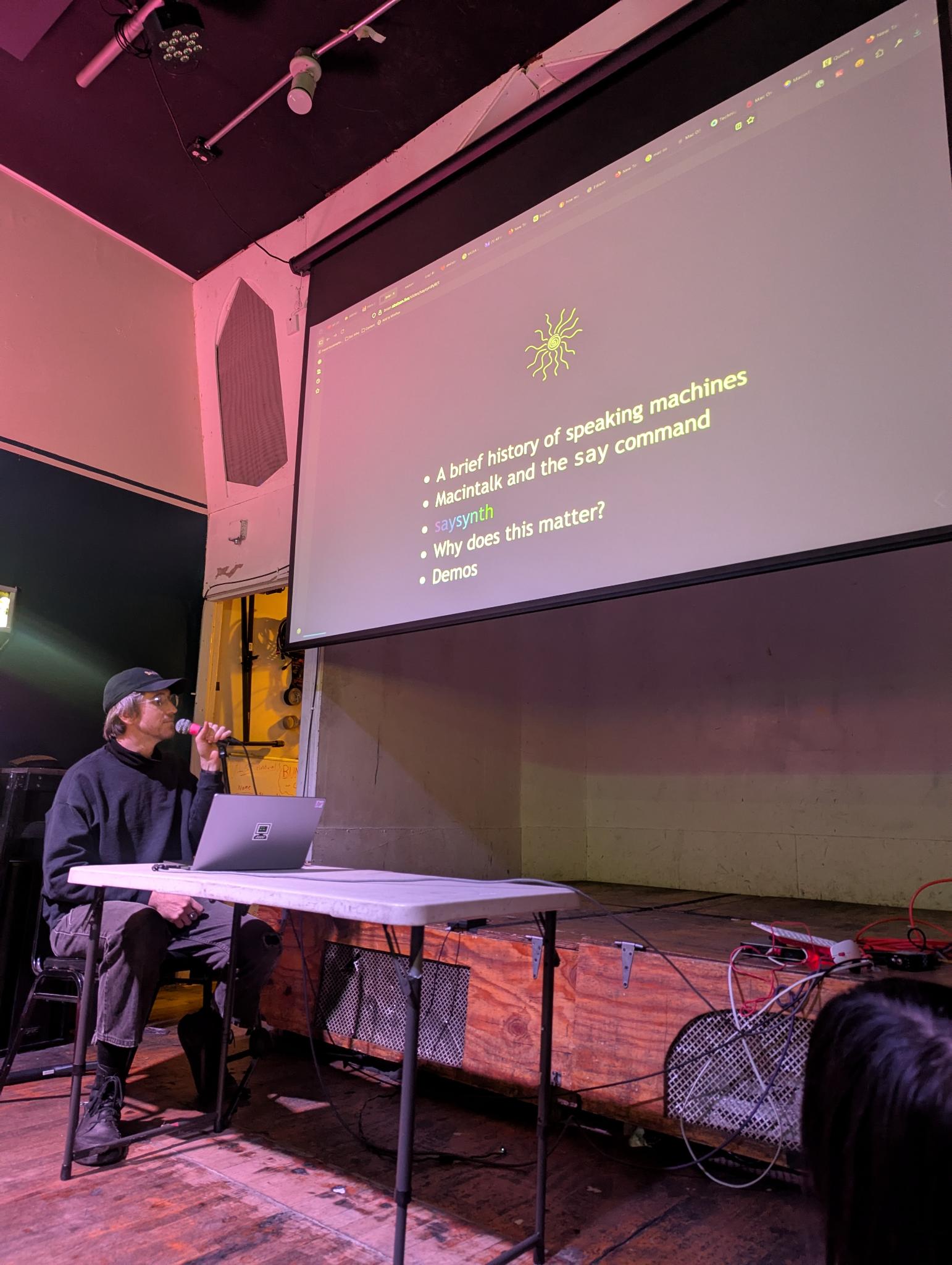}
\caption{Brian Abelson presenting at Composition \#0.}
\Description{Talking into mic in front of a projection of dotted circles and lines. TODO}
\label{fig:appendix-composition-event-brianabelson}
\end{figure}

\begin{figure}[H]
\centering
\includegraphics[width=\linewidth]{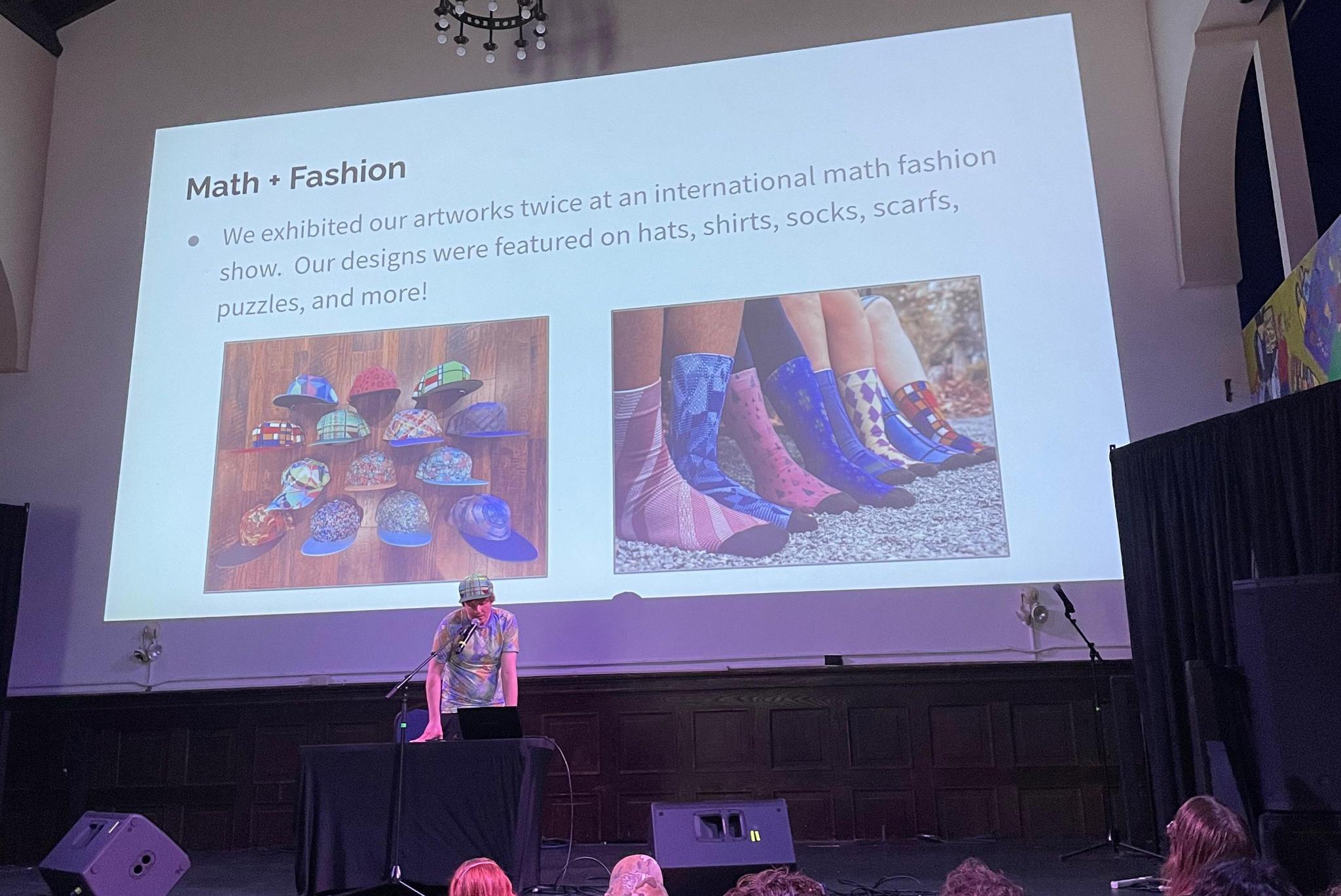}
\caption{Michael Wehar presenting at Composition \#1.}
\Description{Talking into mic in front of a projection of dotted circles and lines. TODO}
\label{fig:appendix-composition-event-michaelwehar}
\end{figure}

\begin{figure}[H]
\centering
\includegraphics[width=\linewidth]{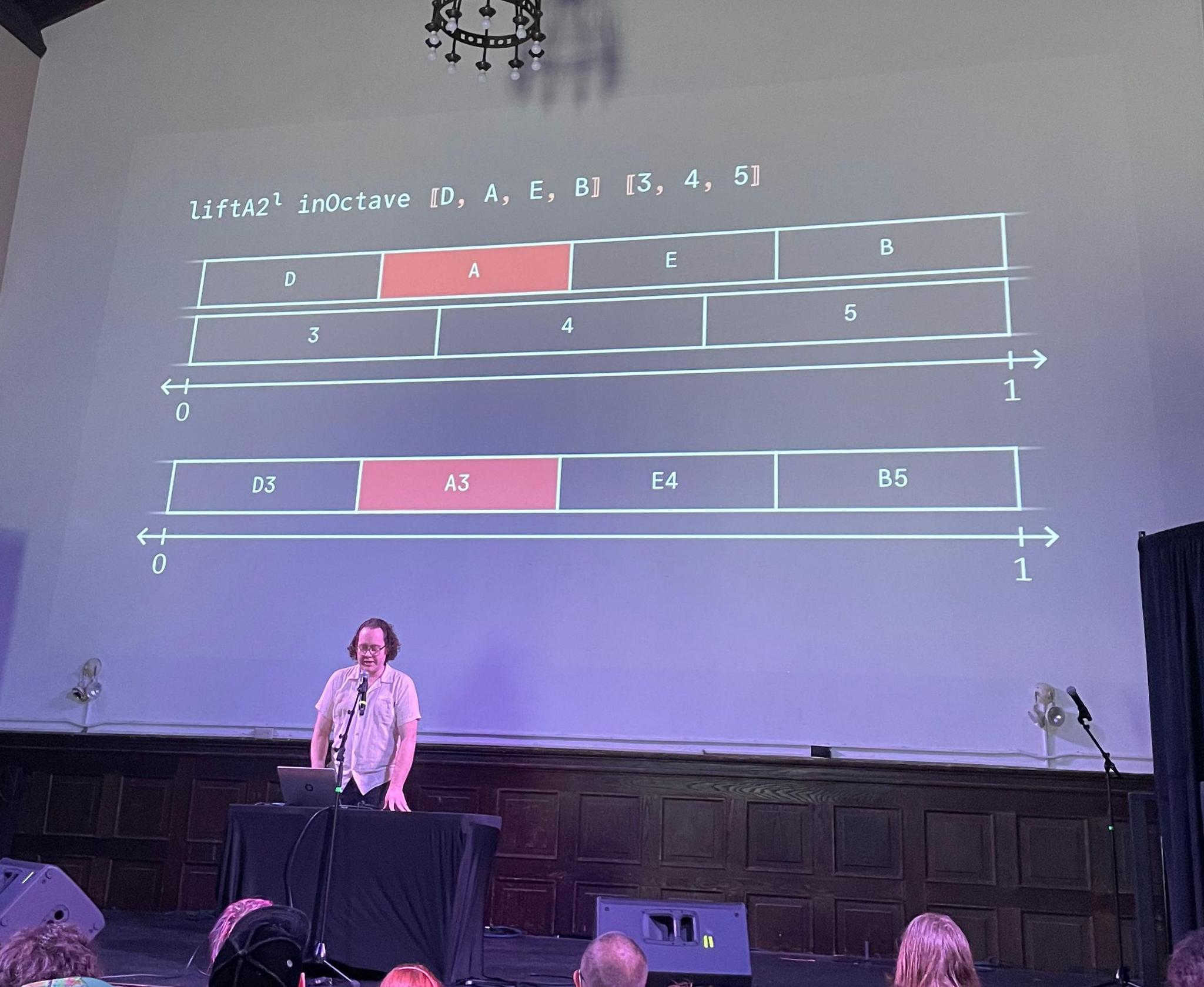}
\caption{Matthew Kaney presenting  at Composition \#1.}
\Description{Talking into mic in front of a projection of dotted circles and lines. TODO}
\label{fig:appendix-composition-event-matthewkaney}
\end{figure}

\begin{figure}[H]
\centering
\includegraphics[width=\linewidth]{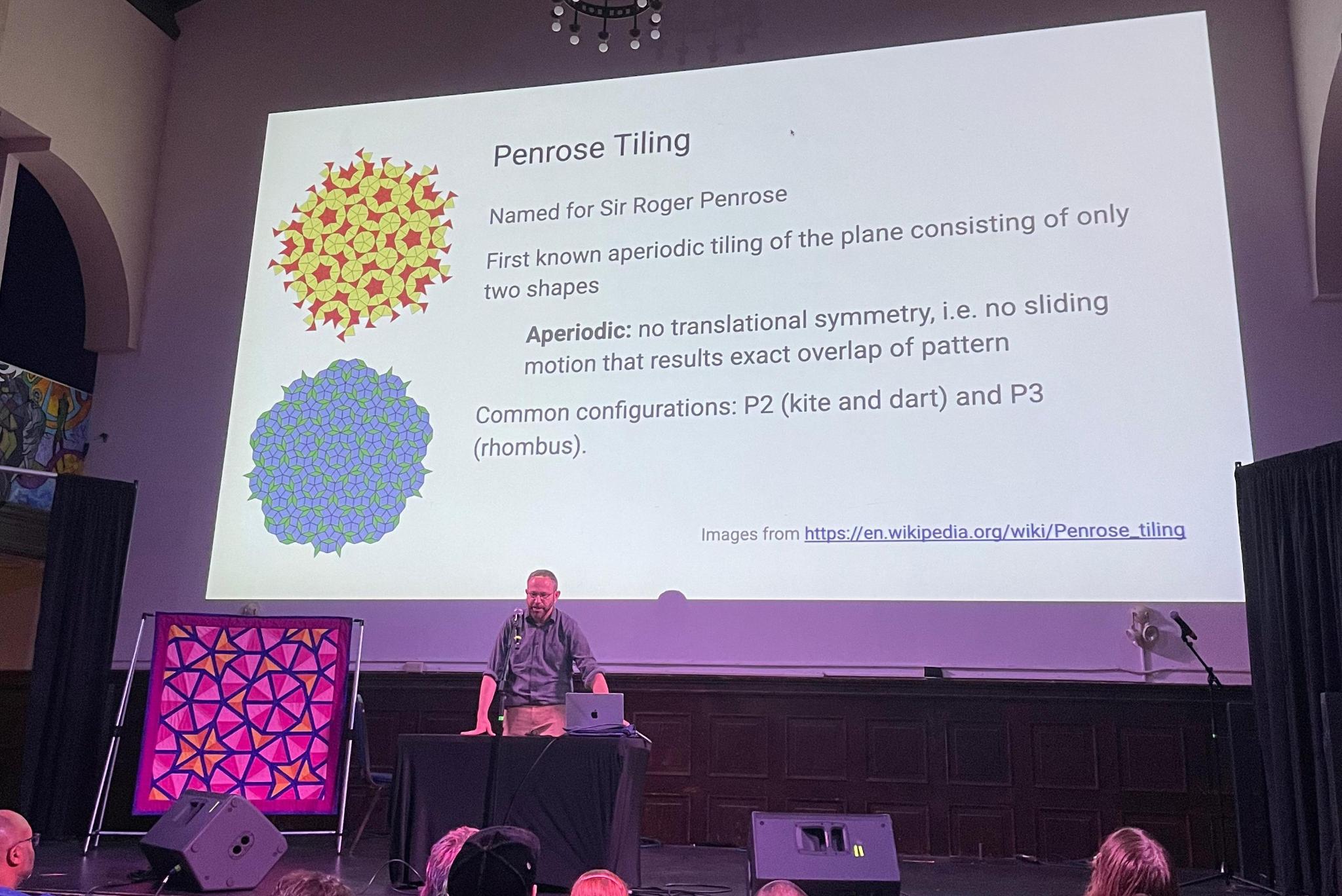}
\caption{Matt Zucker presenting at Composition \#1.}
\Description{Talking into mic in front of a projection of dotted circles and lines. TODO}
\label{fig:appendix-composition-event-mattzucker}
\end{figure}

\begin{figure}[H]
\centering
\includegraphics[width=\linewidth]
{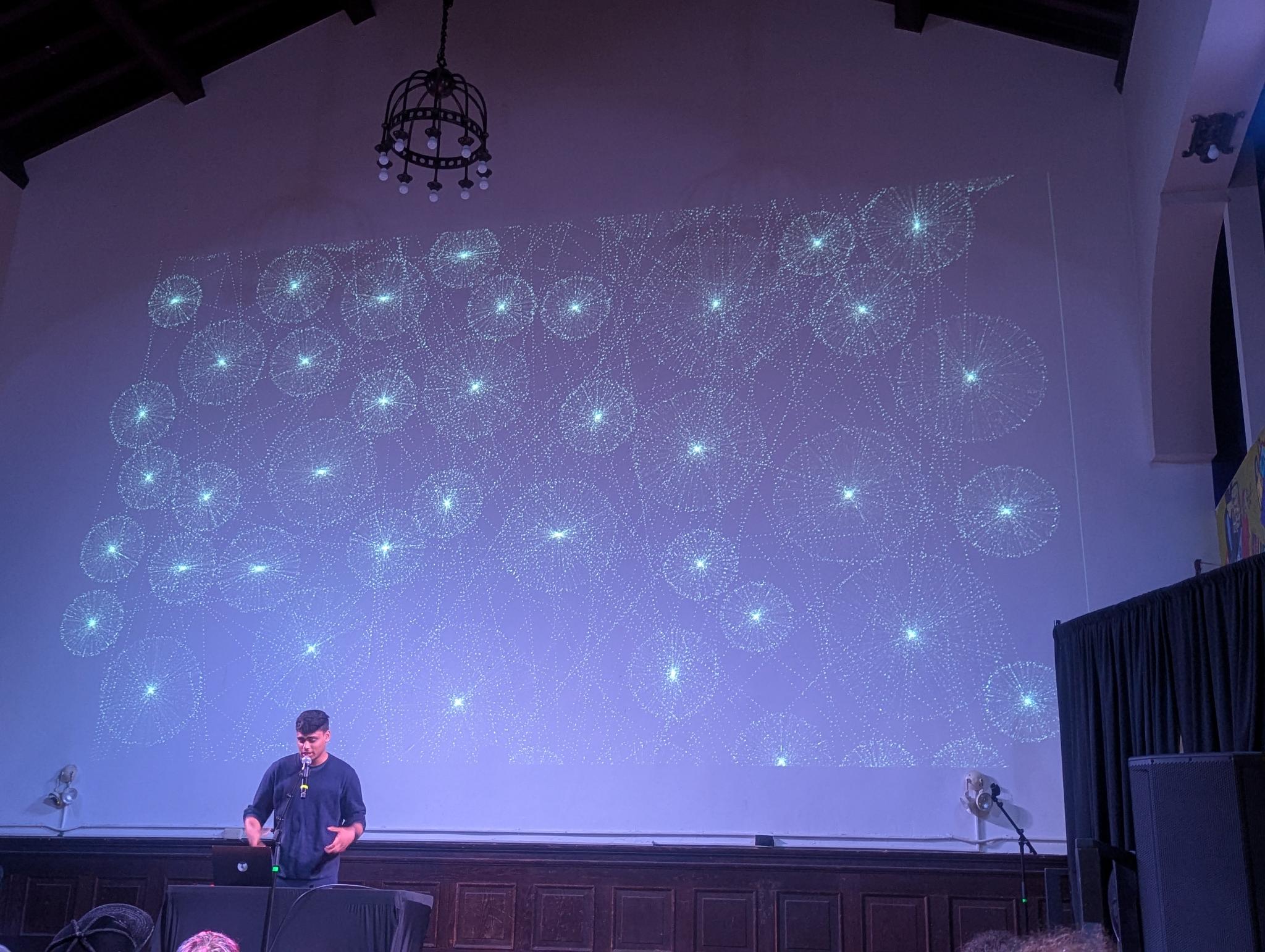}
\caption{arjun presenting at Composition \#1.}
\Description{Talking into mic in front of a projection of dotted circles and lines.}
\label{fig:appendix-composition-event-arjun}
\end{figure}

\begin{figure}[H]
\centering
\includegraphics[width=\linewidth]{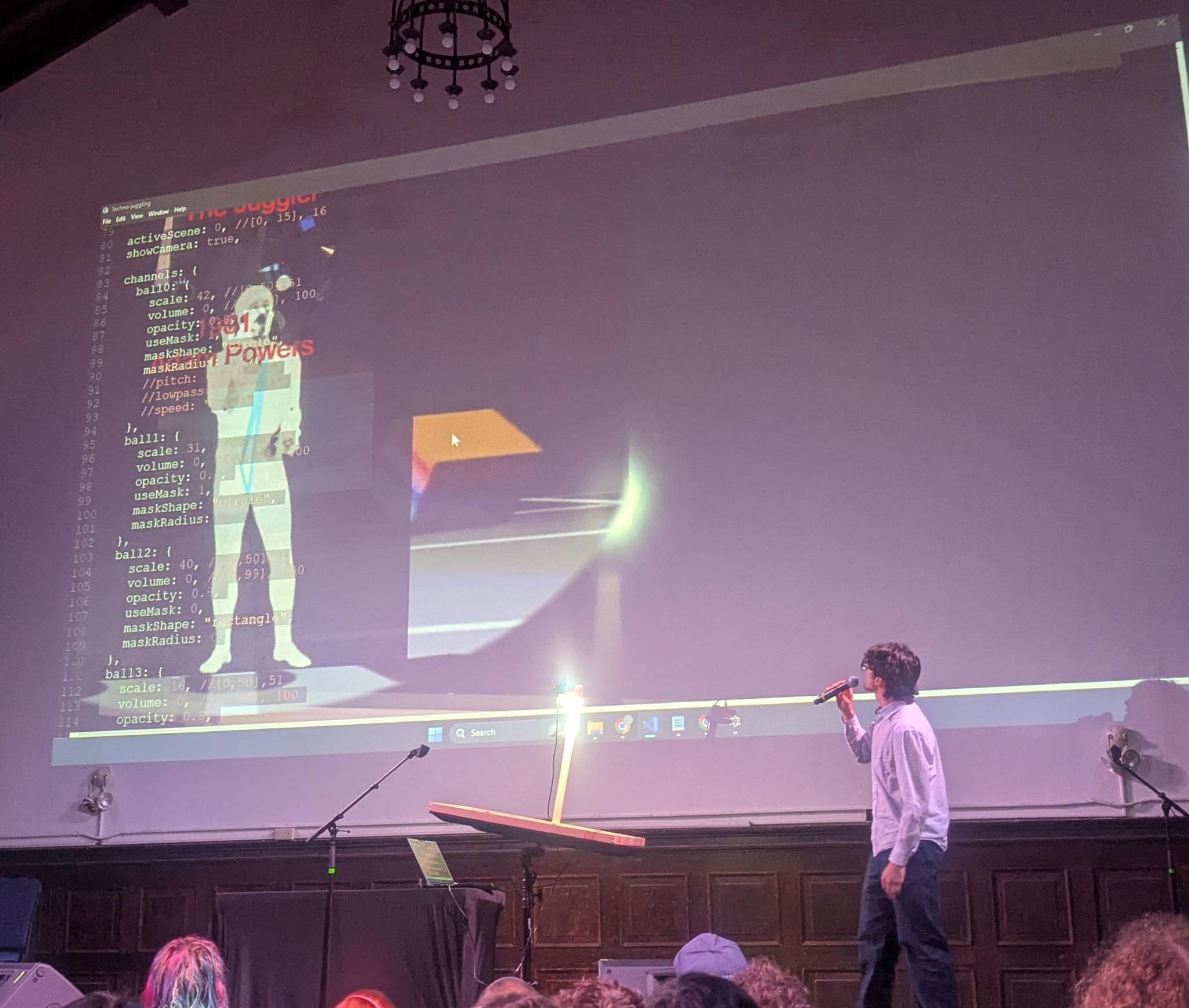}
\caption{Nick Romano presenting at Composition \#1.}
\Description{Talking into mic in front of a projection of dotted circles and lines. TODO}
\label{fig:appendix-composition-event-nickromano}
\end{figure}

%% file: appendix/Graphics.tex
\section{Composition Graphics}
\label{app:graphics}
The following graphics were used to promote the Composition events.


\begin{figure}[H]
    \centering
    \includegraphics[width=0.4\textwidth]{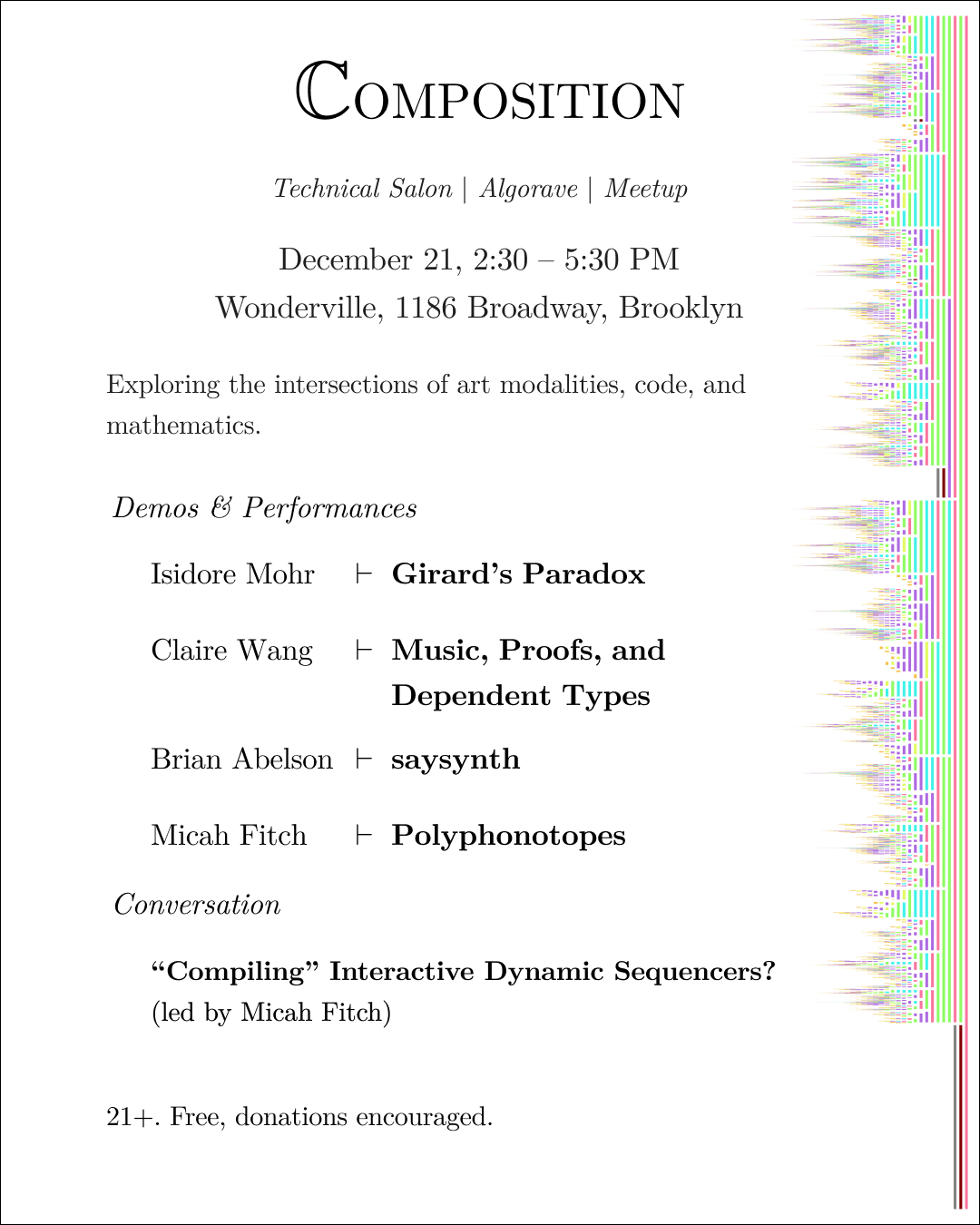}
    \Description{}
    \caption{Event flyer for Composition \#0.}
    \label{fig:flyer0_portrait}
\end{figure}

\begin{figure}[H]
    \centering
    \includegraphics[width=0.4\textwidth]{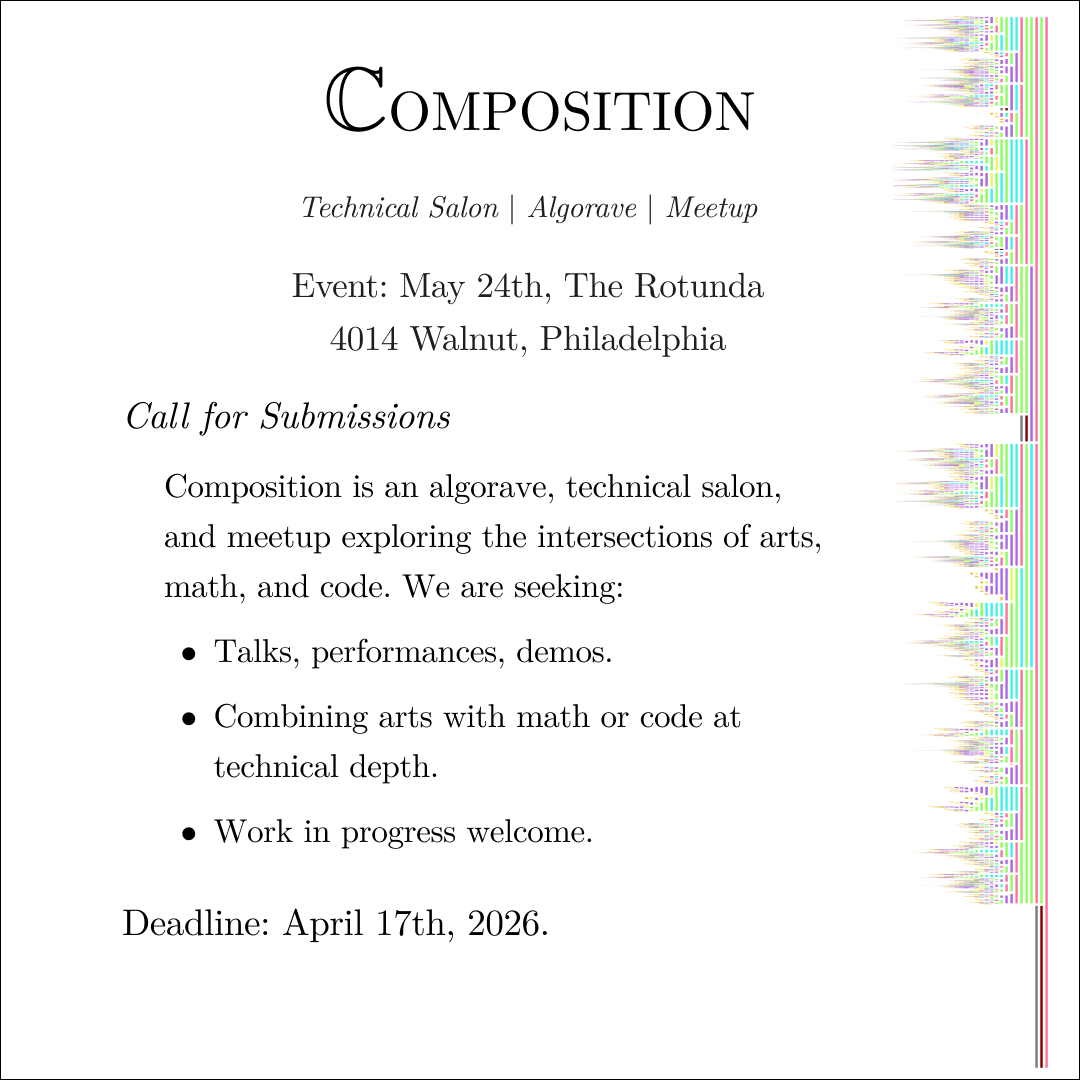}
    \Description{}
    \caption{Submission flyer for Composition \#1.}
    \label{fig:cfs1}
\end{figure}

\begin{figure}[H]
    \centering
    \includegraphics[width=0.4\textwidth]{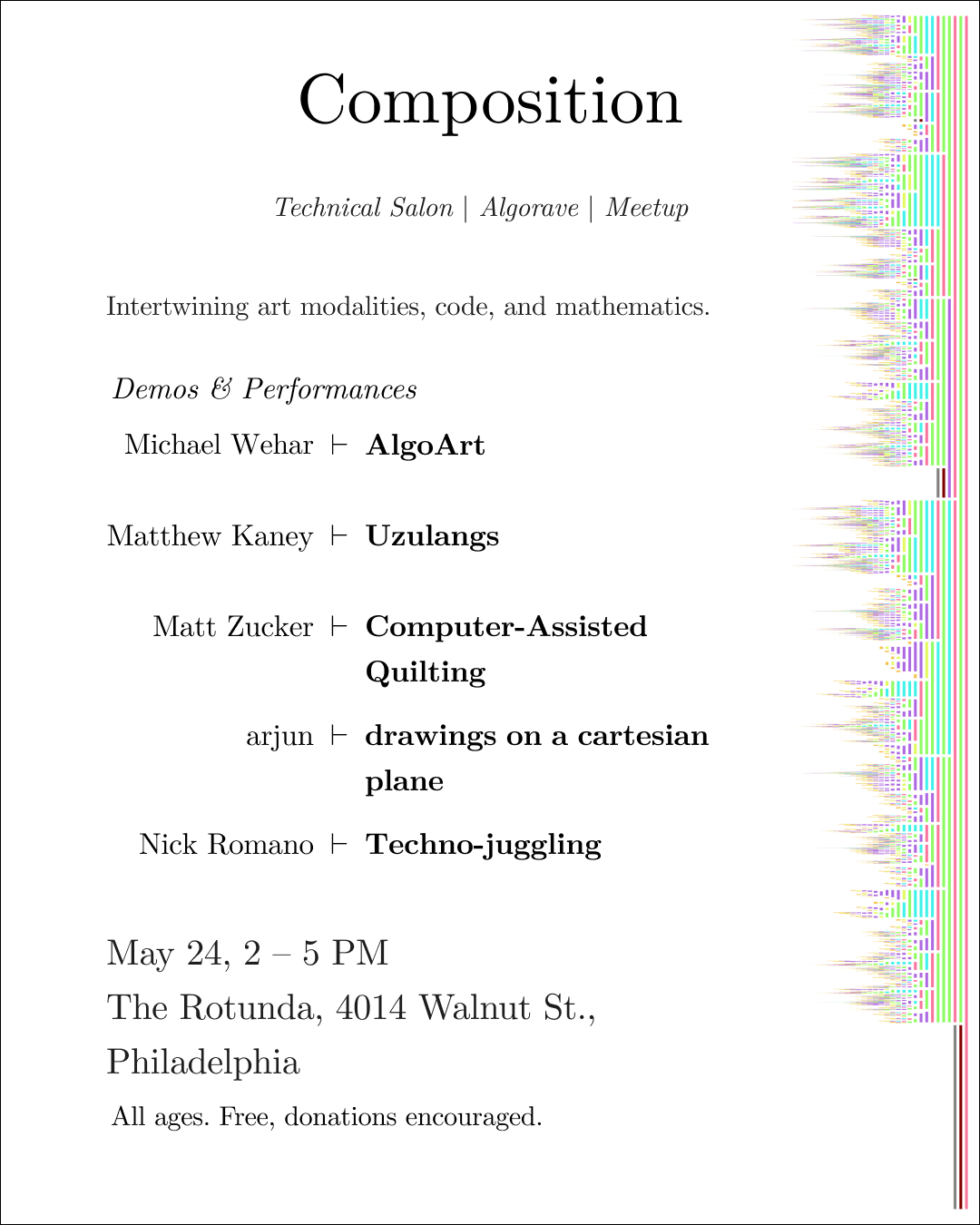}
    \Description{}
    \caption{Event flyer for Composition \#1.}
    \label{fig:flyer1}
\end{figure}

%% file: appendix/CFP.tex
\section{Submissions}
\label{app:submissions}
The following are the questions from the Google form used to collect submissions for Composition \#1. Questions marked with an asterisk are required. The form for Composition \#0 is similar.
\subsection{Submission Form}
\textbf{Email}* \hfill [Short answer] \\[\medskipamount]
\textbf{Name}* \hfill [Short answer] \\[\medskipamount]
\textbf{Socials, Links} \hfill [Short answer] \\[\medskipamount]
\textbf{Proposal guidelines} \\
We're open to anything - a performance, presentation, talk, or workshop, etc. The only ask we have is to show us how your work is related to art, proofs, or programs. Please upload any necessary videos, photos, documents, etc. if needed to convey your submission. \\[\medskipamount]
\textbf{Submission Proposal} \\
Describe what you'll be doing. \hfill [Long answer] \\[\medskipamount]
\textbf{Duration of Submission in minutes}* \hfill [Short answer]\\[\medskipamount]
\textbf{Submission file uploads} \\
Anything to show/add to your submission. \hfill [File upload]\\[\medskipamount]
\textbf{What kind of equipment will you need?} \hfill [Short answer]\\[\medskipamount]
\textbf{Anything else?} \hfill [Long answer]

%% file: appendix/Feedback.tex
\section{Feedback}
\label{app:feedback}
The following are questions from the Composition Feedback Google form, a publicly available Google form for on-going feedback. Questions marked with an asterisk are required.
\subsection{Feedback Form}
\textbf{Which Composition event did you attend?}* \hfill [Multiple choice]
\begin{itemize}
\item 12/21/2025 Composition \#0 at Wonderville, Brooklyn
\item 5/24/2026 Composition \#1 at The Rotunda, Philly
\item I am interested in Composition but haven't attended
\end{itemize}
\textbf{Can you share a few sentences about your experience at the event?}* \hfill [Long answer]\\[\medskipamount]
\textbf{What could be improved?} \hfill [Long answer]\\[\medskipamount]
\textbf{What did you like?} \hfill [Long answer]\\[\medskipamount]
\textbf{Do you have any other suggestions or comments?} \hfill [Long answer] \\[\medskipamount]
\textbf{What did you like?} \hfill [Long answer]